\documentclass[prl,twocolumn,a4paper]{revtex4}
\usepackage{amssymb}
\usepackage{epsfig}
\usepackage{amsmath}
\usepackage{graphicx}
\usepackage{bm}
\usepackage{times}
\usepackage{color} 
\usepackage{ulem}

\begin{document}

\title{Strong coupling of a mechanical oscillator and a single atom}

\author{K. Hammerer$^{1,5}$, M. Wallquist$^{1,5}$, C. Genes$^{1}$, M. Ludwig$^2$, F. Marquardt$^2$, P. Treutlein$^3$, P. Zoller$^{1,5}$, J. Ye$^{4,5}$, H. J. Kimble$^5$}

\affiliation{
$^1$ \mbox{Institute for Theoretical Physics, University of Innsbruck, and Institute
for Quantum Optics and Quantum Information,} Austrian Academy of Sciences,
Technikerstrasse 25, 6020 Innsbruck, Austria\\
$^2$ Department of Physics, Center for NanoScience, and Arnold Sommerfeld
Center for Theoretical Physics, Ludwig-Maximilians-Universit\"at M\"unchen,
Theresienstr. 37, D-80333 Munich, Germany\\
$^3$ Max-Planck-Institute of Quantum Optics and Department of Physics, Ludwig-Maximilians-Universit\"at M\"unchen, Schellingstr. 4, D-80799 Munich, Germany\\
$^4$ JILA, National Institute of Standards and Technology and University of Colorado, Boulder, CO 80309-0440 USA\\
$^5$ Norman Bridge Laboratory of Physics 12-33, California Institute of Technology, Pasadena, CA 91125 USA}

\date{\today}
\begin{abstract}
We propose and analyze a setup to achieve strong coupling between a single trapped atom and a mechanical oscillator. The interaction between the motion of the atom and the mechanical oscillator is mediated by a quantized light field in a laser driven high-finesse cavity. In particular, we show that high fidelity transfer of quantum states between the atom and the mechanical oscillator is in reach for existing or near future experimental parameters. Our setup provides the basic toolbox for coherent manipulation, preparation and measurement of micro- and nanomechanical oscillators via the tools of atomic physics.
\end{abstract}
\maketitle

\begin{figure}[t]
\includegraphics[width=0.8\columnwidth]{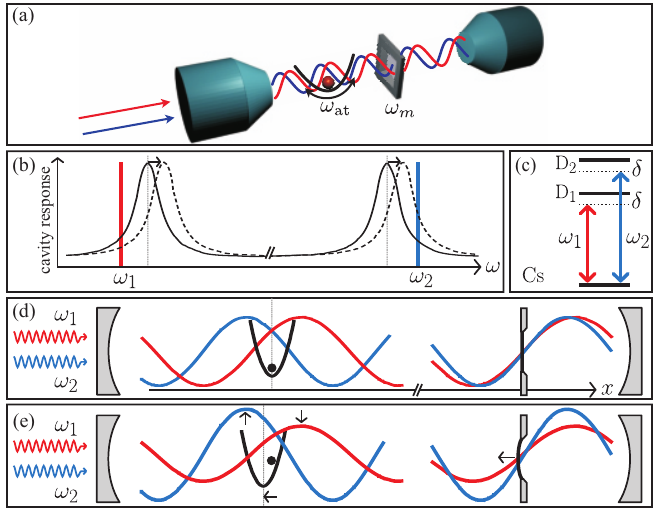}
\caption{(a) Strong coupling of the motion of a single atom to a vibrational degree
of freedom of a micron-sized membrane can be achieved in a two mode cavity (for details see text). (b) Cavity response as a function of frequency. Two cavity modes are driven by two lasers of frequencies $\omega_{1}$ and $\omega_{2}$, with red and blue detuning respectively. (c) The two frequencies drive two atomic transitions, e.g. the D$_{1,2}$ lines of Cs, both with red detuning, causing AC Stark shift of the ground state. (d) (left side) The atom is trapped in the potential from the two optical lattices (red and blue curves) $u_{1,2}(x)=\sin^{2}\left(k_{1,2}x\right)$ with wave vectors $k_{1}\ne k_{2}$. (right side) The membrane is placed at a point of steepest slope of the intensity profiles $u_{1,2}(x)$ where opto-mechanical coupling is maximal. (e) A small displacement of the membrane will shift the cavity resonances [cf. dashed line in (b)] resulting in a spatial shift of the trap potential for the atom, and thus an effective linear atom-membrane coupling as in Eq.~\eqref{Eq:Heff}. (Displacements and frequency shifts are not to scale.)}
\label{Fig:Coupling}
\end{figure}

Recent experiments with micro- and nanomechanical oscillators coupled to the optical field in a cavity are approaching the regime
where quantum effects dominate \cite{Schliesser2009,Groeblacher2009,Marquardt2009}. In light of this progress, the question arises to what extent the quantized motion of a mesoscopic mechanical
system can be coherently coupled to a microscopic quantum object \cite{Tian2004,Treutlein2007,Lambert2008,Rabl2009,Genes2008,Singh2008},
the ultimate challenge being strong coupling to the motion of a single
atom. For a direct mechanical coupling the interaction involves scale
factors $\sqrt{m/M}\sim10^{-7}-10^{-4}$ depending on the ratio of the
mass of the atom $m$ to the mass of the mechanical oscillator $M$ \cite{Tian2004}. It is therefore difficult to achieve a coherent coupling for exchange of a single vibrational
quantum that is much larger than relevant dissipation rates.

In this Letter we show, however, that strong coupling can be realized between a single trapped atom and an opto-mechanical oscillator.  The coupling between the motion
of a membrane \cite{Thompson2008} -- representing the mechanical oscillator -- and the
atom is mediated by the quantized light field in a laser driven high-finesse
cavity. Remarkably, in this setup a coherent coupling for single-atom
and membrane exceeding the dissipative rates by a factor
of ten is within reach for present or near future experimental parameters \cite{Miller2005}.
Entering the strong coupling regime provides a quantum interface
allowing the coherent transfer of quantum states 
between the mechanical oscillator and atoms, opening the door to coherent manipulation, preparation and measurement of micromechanical
objects via the well-developed tools of atomic physics.

We propose and analyze a setup which combines the recent advances
of micromechanics with membranes in optical cavities \cite{Thompson2008} and cavity QED with single trapped atoms \cite{Miller2005} (see Fig.~\ref{Fig:Coupling}a). We consider
a membrane placed in a laser driven high-finesse cavity representing the
opto-mechanical system with radiation pressure coupling.
In this setup the motion of the membrane manifests itself as a dynamic detuning
of cavity modes. For a cavity mode driven by a detuned laser
this translates into a variation of the intensity of the intracavity
light field. In addition, we assume that this intracavity field provides
an optical lattice as a trap for a single atom. Thus for the setup
of Fig.~\ref{Fig:Coupling}a the motion of the membrane will be coupled
via the dynamics of the optical trap to the motion of the atom, and vice versa. This
coupling is strongly enhanced by the cavity finesse which is a key
ingredient in achieving the strong coupling regime.

In the following we are interested in a configuration which - after
integrating out the internal cavity dynamics - realizes a coupled oscillator dynamics
\textit{linear }in the displacements of atom and membrane $(\hbar=1)$
\begin{equation}\label{Eq:Heff}
H=\omega_{m}a_{m}^{\dagger}a_{m}^{\phantom{\dagger}}
+\omega_{at}a_{at}^{\dagger}a_{at}^{\phantom{\dagger}}
-G(a_{at}+a_{at}^{\dagger})(a_{m}+a_{m}^{\dagger}).
\end{equation}
The first and second term are the Hamiltonians of the bare micromechanical
oscillator and the harmonic motion of the trapped
atom, respectively. We adopt the notation
$x_{\mu}\equiv\ell_{\mu}(a_{\mu}+a_{\mu}^{\dagger})$
and $p_{\mu}$ for the position and momentum operators (along the cavity axis)
with $\mu\equiv (m,at)$ for
the membrane and atom, respectively, and $a_{\mu}$ are annihilation
operators.
Both atom and mechanical oscillator are prepared close to their respective ground states, and their oscillator lengths are denoted by
$\ell_{m}=\sqrt{\hbar/2M\omega_{m}}$ and $\ell_{at}=\sqrt{\hbar/2m\omega_{at}}$
with $\ell_{m}\ll\ell_{at}$ in view of $M\gg m$, and we assume a
near resonance condition $\omega_{m}\approx\omega_{at}$
of the mechanical and atomic oscillation frequencies. The system dynamics will obey a master equation
\begin{equation}\label{Eq:EffMEQ}
\dot{\rho}=-i\left[H,\rho\right]+(L_{c}+L_{at}+L_{m})\rho,
\end{equation}
where the three Liouvillian terms describe dissipation via cavity decay, atomic momentum diffusion due to spontaneous emission, and thermal heating of the membrane, respectively. Our goal is to obtain a coupling $G$ much larger than the rates of decoherence through these channels.

A strong effective coupling as in Eq.~(\ref{Eq:Heff})
is obtained in a configuration involving two cavity modes (Fig.~\ref{Fig:Coupling}).
The two modes are driven by lasers of frequencies $\omega_{1}$ and
$\omega_{2}$, respectively, where the first (second) laser is tuned
to the red (blue) side of its respective cavity resonance (Fig.~\ref{Fig:Coupling}b,c).
Both lasers provide red-detuned optical lattices for the atom with wave vectors $k_1 \neq k_2$. A single atom is trapped in one of the wells of the combined potential of the two lattices (Fig. 2d). The particular well within the optical lattice array is chosen such that
each of the two potentials has close to maximal but opposite slope at the equilibrium
position $\bar{x}_{at}$ of the atom. The membrane in turn is positioned
at $\bar{x}_{m}$ half-way between a field node and anti-node, with similar slope for both modes, where
the linear opto-mechanical coupling is maximal \cite{Thompson2008}. A small displacement
of the membrane will thus shift the cavity resonances {[}cf. dashed
line in Fig.~\ref{Fig:Coupling}b{]}. Accordingly, one driving laser
will come closer to resonance, the other one farther
off resonance. This will in turn make one of the lattice potentials
deeper, the other one shallower, giving rise to a \textit{spatial shift} of
the atomic trapping potential proportional to $x_{m}$ (Fig.~\ref{Fig:Coupling}e),
resulting in an overall $\sim x_{at}x_{m}$ coupling as in Eq.~\eqref{Eq:Heff}.

Before we analyze this setup in detail we note that for a \textit{single} standing-wave cavity mode a displacement of the membrane $x_{m}$ results in a change of the potential \textit{depth} and thus a \textit{parametric} coupling of the atom to the motion of the membrane of the type $\sim x_{m}x_{at}^{2}$.
This parametric coupling, which is in principle present also in the proposed two mode setup, will be smaller than the linear coupling in Eq.~\eqref{Eq:Heff} by at least a Lamb-Dicke factor $\eta=k_1\ell_{at}\ll1$ and can be neglected here. In the following we will first explain the coupling of the two cavity modes to displacements of the atom and the membrane, including the relevant decay mechanisms. In the second step we adiabatically eliminate the cavity mode and derive the effective system dynamics as given by Eq.~\eqref{Eq:EffMEQ}. This will allow us to identify the requirements for strong coupling.

\textit{Atom--cavity interaction:} The optical potential along
the cavity axis seen by the atom is $V(x)= U_{0}\big(u_{1}(x)A_{1}^{\dagger}A_{1}^{\phantom{\dagger}}+u_{2}(x)A_{2}^{\dagger}A_{2}^{\phantom{\dagger}}\big),$
where $u_{i}(x)=\sin^2(k_{i}x)$ and $A_{i}$ is a photon destruction
operator for field modes $i=1,2$. We assume for simplicity that each of
the cavity fields generates the same AC Stark shift $U_{0}=\frac{\Omega_{0}^2}{\delta}$
per photon, where $\Omega_{0}$ is the vacuum Rabi frequency and $\delta<0$
is the detuning from atomic resonance (see Fig.~\ref{Fig:Coupling}).
In our effective 1D model, transverse confinement is naturally provided by the Gaussian intensity
profile of the cavity fields. Consider the case where both cavity
fields are driven so that we have a large intracavity amplitude
$\alpha$, which we choose to be equal and real for both cavity modes. Expanding the potential in powers of this amplitude yields $V(x)\simeq U_{0}\alpha^{2}u(x)+ U_{0}\alpha\big[u_{1}(x)a_{1}+u_{2}(x)a_{2}+\mathrm{h.c.}\big]$,
where $u(x)=u_{1}(x)+u_{2}(x)$, and we neglected terms of order zero
in $\alpha$. The operators $a_i$ describe amplitude fluctuations around the coherent field $\alpha$, i.e. $A_i=\alpha+a_i$. The first term $\sim u(x)$ is the effective atomic potential created by the combined effect of the two cavity modes.

In a Lamb-Dicke expansion around the equilibrium position $\bar{x}_{at}$, the potential together with the kinetic energy of the atom combine to $p^{2}/2m+V(x)\rightarrow\omega_{at}a_{at}^{\dagger}a_{at}^{\phantom{\dagger}}+H_{at,c}$,
where \begin{equation}\label{Eq:Hatc}
H_{at,c}=g_{at,c}\big[(a_{1}+a_{1}^{\dagger})-(a_{2}+a_{2}^{\dagger})\big]
(a_{at}^{\phantom{\dagger}}+a_{at}^{\dagger}),
\end{equation}
and we adopt for the motion of the atom a harmonic approximation with
a trap frequency $\omega_{at}^{2}=U_{0}\alpha^{2}u''(\bar{x}_{at})/m$.
Here $H_{at,c}$ is the desired linear atom-field coupling at rate $g_{at,c}=U_{0}\alpha\,\eta\,\theta$, where $\theta=\frac{u_{1}'(\bar{x}_{at})}{k_{1}}$ is a geometrical factor. We assume that the $\bar{x}_{at}$ is chosen such that $\theta\simeq1$. This interaction can be interpreted
as follows: Fluctuations in the amplitudes of the two cavity fields,
as quantified by the quadrature operators $a_{i}+a_{i}^{\dagger}$,
exert \textit{oppositely oriented forces} on the atom. Conversely, fluctuations
of the atom around its mean position, as quantified by $a_{at}^{\phantom{\dagger}}+a_{at}^{\dagger}$,
cause changes of \textit{opposite sign} in the amplitudes of the two
cavity fields.

\textit{Membrane--cavity interaction:} As demonstrated
\cite{Thompson2008}, vibrational fluctuations of a thin dielectric
membrane couple to cavity quadratures according to
\begin{equation*}
H_{m,c}=g_{m,c}\big[(a_{1}+a_{1}^{\dagger})+(a_{2}+a_{2}^{\dagger})\big]
(a_{m}^{\phantom{\dagger}}+a_{m}^{\dagger}),
\end{equation*}
with an opto-mechanical coupling $g_{m,c}=\frac{\ell_{m}}{L}\omega_i f_{i}(\bar{x}_{m})\alpha~(i=1,2)$, which we take for simplicity to be the same for both cavity fields. $L$ is the length of the cavity.
The geometrical factor $f_{i}(\bar{x}_{m})=2r\sin(2k_{i}\bar{x}_{m})/\sqrt{1-r^{2}\cos^2(2k_{i}\bar{x}_{m})}$
depends on the membrane amplitude reflectivity $r$ and the equilibrium
position $\bar{x}_{m}$ of the membrane. By a proper choice of $\bar{x}_{m}$
it is possible to achieve $f_{i}\simeq2r$ for both fields. The interpretation of this coupling is completely analogous to the one of the atom-cavity interaction in Eq.~\eqref{Eq:Hatc}.

\textit{Open system dynamics:} For the combined system
of Fig.~\ref{Fig:Coupling}a we thus arrive at a Hamiltonian
\[
H=\omega_{at}a_{at}^{\dagger}a_{at}^{\phantom{\dagger}}
+\omega_{m}a_{m}^{\dagger}a_{m}^{\phantom{\dagger}}
-\Delta (a_{1}^{\dagger}a_{1}^{\phantom{\dagger}}-a_{2}^{\dagger}a_{2}^{\phantom{\dagger}})
+H_{at,c}+H_{m,c}.
\]
For the two cavity fields this Hamiltonian refers to frames rotating
at the respective driving laser frequencies $\omega_{i}$, see Fig.~\ref{Fig:Coupling}. The laser detunings,
$\pm\Delta$, for the two cavity modes are chosen equal in magnitude and opposite in sign. The coherent evolution described by this Hamiltonian is accompanied by various decay channels, such that the density matrix $W$ of the entire system comprising the atom, the membrane and the two cavity fields evolves according to a master equation $\dot W=-i[H,W]+(L_{1}+L_{2}+L_{at}+L_m)W$. Using
the notation $D[a]W=2aW a^{\dagger}-a^{\dagger}aW-W a^{\dagger}a$
to denote a general Lindblad term, we have in particular $L_{1,2}W=\kappa D[a_{1,2}]W$
with a cavity amplitude decay rate $\kappa$. Spontaneous emission will
inevitably cause momentum diffusion of the atom, which is described by $L_{at}W=\frac{\Gamma_{at}}{2}D[a_{at}+a_{at}^{\dagger}]W$
and happens at a rate $\Gamma_{at}=\gamma\frac{\Omega_{0}^{2}\alpha^{2}}{\delta^{2}}\eta^{2}u(\bar{x}_{at})
=\gamma\frac{g_{at,c}^{2}}{\Omega_{0}^{2}}\,\xi$, where
$\gamma$ is the spontaneous decay rate \footnote{We assume that losses
to other levels can be excluded.}. The geometrical factor $\xi=\frac{k_{1}^{2}u(\bar{x}_{at})}{{u'_1}(\bar{x}_{at})^2}$ can be made close to unity by a proper choice of $\bar{x}_{at}$ \footnote{We require $\xi(\bar{x}_{at}),\,\theta(\bar{x}_{at})\simeq 1$. For the two cavity modes with wave numbers $k_{1,2}=\frac{1}{2}(k\pm\delta k)$ the intensity extrema fulfil $k\tan(k\, x)=-\delta k\tan(\delta k\, x)$. The potential minima close to points where $\delta k\, x\simeq n\pi$ have the desired properties.}. Finally, thermal contact of the membrane to the environment
at a temperature $T$ is accounted for by $L_{m}W=\frac{\gamma_m}{2}(\bar{n}+1)D[a_{m}]W+\frac{\gamma_m}{2}\bar{n}D[a_{m}^{\dagger}]W$,
where $\gamma_{m}$ is the natural linewidth of the mechanical resonance
and $\bar{n}$ its mean occupation in thermal equilibrium. The relevant
effective decoherence rate of the membrane is $\Gamma_{m}=\gamma_m\bar{n}\simeq\frac{k_{B}T}{\hbar Q}$ for a mechanical quality factor $Q$.

\textit{Mediated atom-membrane interaction:} We are now in the position to derive the effective cavity--mediated coupling between the single atom and the membrane.
Consider the case of far off-resonant drive $|\Delta|\gg g_{at,c},g_{m,c}$,
where fluctuations in cavity quadratures are fast variables and adiabatically
follow the dynamics of
position fluctuations of atom
and membrane. In this dispersive limit the decoherence rate due to cavity decay can be kept small as compared to the strength of coherent evolution by choosing $\frac{\kappa}{\Delta}\ll1$. We derive an effective master equation for the reduced state of atom and membrane $\rho={\rm tr}_{12}\{W\}$ as given in Eq.~\eqref{Eq:EffMEQ}. The rate of mediated coherent coupling described by the Hamiltonian in Eq.~\eqref{Eq:Heff} is
\[
G=\frac{2g_{at,c}g_{m,c}(\Delta+\omega_{m})}{\kappa^{2}+(\Delta+\omega_{m})^{2}}
+\frac{2g_{at,c}g_{m,c}(\Delta-\omega_{m})}{\kappa^{2}+(\Delta-\omega_{m})^{2}}.
\]
The most compelling feature of this cavity mediated ``spring'' is that -- to the best of our knowledge -- this is the first scheme for coupling the motion of a single atom to a massive oscillator which manages to avoid the mass ratio $\sqrt{m/M}$ entering the coupling strength. This ratio necessarily enters any translationally invariant coupling $\sim(x_{at} - x_m)^2$, as it sets the relative magnitude of the cross-term $\sim x_{at} x_m$ versus the direct atomic frequency shift term $\sim x_{at}^2$.

Decay of the cavity field gives rise to four channels of decoherence in the effective master equation in Eq.~\eqref{Eq:EffMEQ},
\begin{equation}
L_{c}\rho =\sum_{\sigma=\pm}\frac{\Gamma_{c}^{\sigma}}{2}D[J_{\sigma}]\rho
+\frac{\Gamma_{c}^{-\sigma}}{2}D[J_{\sigma}^{\dagger}]\rho
\label{Eq:Lc}
\end{equation}
at rates $\Gamma_{c}^{\pm}=\frac{2\kappa(g_{at,c}^{2}+g_{m,c}^{2})}{\kappa^{2}+(\Delta\pm\omega_{m})^{2}}$
with jump operators $J_{\pm}=\cos(\phi)\, a_{m}\pm\sin(\phi)\, a_{at}$
where $\tan\phi=\frac{g_{at,c}}{g_{m,c}}$. Each
of the four decay channels contributing to $L_{c}\rho$ is associated
with emission of sideband photons at either side of the two driving
lasers, that is, at one of the frequencies $\omega_{1,2}\pm\omega_{m}$.
An emission event is accompanied by the creation or annihilation of
a quantum in either atom or membrane. For a near resonant
system ($\omega_{m}\simeq\omega_{at}$) these two possibilities are indistinguishable,
such that both processes happen in a coherent fashion. Therefore,
the jump operators $J_{\pm}$ are linear combinations of the corresponding
creation/annihilation operators.

\textit{Strong coupling regime:} We now show that the coupling can be strong enough such that coherent dynamics dominates over the various decoherence processes. In a system described by the effective
master equation \eqref{Eq:EffMEQ} strong coupling is established
by fulfilling the set of conditions
\begin{equation}
G \gg\Gamma_{c}^{\pm},\Gamma_{at},\Gamma_{m},\label{Eq:Cond0}
\end{equation}
in addition to $\omega_{at}=\omega_{m}$ for a resonant coupling. For a ratio $\frac{\Gamma_{c}^{\pm}}{G}\ll 1$, it is necessary to drive the cavity far off-resonant
\begin{equation}
\Delta \gg \kappa,\omega_m,\label{Eq:Cond1}
\end{equation}
and it is desirable to keep at the same time a balanced atom--cavity and membrane--cavity
coupling $g_{at,c}\simeq g_{m,c}$, which is equivalent to
\begin{equation}\label{Eq:Cond2}
\frac{4r}{\pi}\frac{\delta}{\gamma}\frac{\mathcal{F}}{C}\sqrt{\frac{m}{M}}\simeq1,
\end{equation}
where $C=\frac{\Omega_{0}^{2}}{\kappa\gamma}$ is the 1--atom cooperativity parameter and
$\mathcal{F}=\frac{\pi c}{2\kappa L}$ the cavity finesse. Small decoherence due to atomic momentum diffusion, $\frac{\Gamma_{at}}{G}\ll1$, requires a large cooperativity parameter
\begin{equation}
C\gg\frac{\Delta}{4\kappa}.\label{Eq:Cond3}
\end{equation}
Finally, thermal decoherence depends on the ambient temperature $T$
of the membrane. It is important to note that there is a natural lower limit for the temperature $T$ which is set by light absorption inside the membrane. If we assume the cavity finesse to be limited by absorption, the power absorbed by the membrane is $P_{a}\simeq\frac{2\pi}{\mathcal{F}}P_{c}$
for an overall circulating power $P_{c}=\frac{\hbar\omega_1c\alpha^{2}}{L}$
in the two cavity modes. Such an amount of absorbed power will
cause an increase of the membrane temperature $\Delta T\simeq\frac{1}{k_{B}\kappa_{th}}P_{a}$,
where $\kappa_{th}$ is the thermal link of the membrane to its supporting frame which depends
on the specific geometry and material properties \footnote{$\kappa_{th}$ is
chosen here such as to have dimensions of Hz.}. While it is not entirely
clear how this heating exactly affects the vibrational mode in question, a safe assumption is an equal increase in temperature. For typical parameters (see below), $\Delta T$ corresponds to a few Kelvin, so that standard cryogenic precooling allows one to reach $T\simeq\Delta T$. Under these fairly cautious assumptions we
can expect a small thermal decoherence $\frac{\Gamma_{m}}{G}\ll1$ as long as
\begin{equation}\label{Eq:Cond4}
\frac{8r^{2}}{\pi^{2}}\frac{\kappa_{th}}{\gamma_{m}}\frac{\hbar\omega_1}{Mc^{2}}\mathcal{F}^{2}
\gg\frac{\Delta}{\kappa}.
\end{equation}
Remarkably, this is independent of circulating power and only implicitly depends on temperature through $\kappa_{th}$ \cite{Zink2004}.

Together, Eqs.~\eqref{Eq:Cond1} to \eqref{Eq:Cond4} ensure the set of conditions for strong coupling in \eqref{Eq:Cond0}. Note that the intracavity amplitude $\alpha$ and therefore
the absolute timescale of the system are not fixed by Eqs.~\eqref{Eq:Cond1}
to \eqref{Eq:Cond4}. These equations actually impose conditions on
the properties of the system at the single photon level. The necessary
cavity amplitude $\alpha$, and with it the absolute timescale of
the dynamics, will finally follow from the resonance condition $\omega_{at}=\omega_{m}$.

\textit{Example:} We will show now that the interaction between a single Cs atom and a SiN membrane of small effective mass $M=0.4$\,ng mediated by a high-finesse optical micro--cavity can enter the strong coupling regime. Firstly, we assume a large cavity finesse of $\mathcal{F}\simeq 2\times 10^5$ which is consistent with a measured value of ${\rm Im}(n)\simeq1\times10^{-5}$ for the absorption in a
SiN membrane inside a cavity \cite{Wilson2009}. A small cavity waist of $w_0=10~\mu$m results in a cooperativity parameter of $C=140$. A ratio of $\frac{\Delta}{\kappa}\simeq 18$ satisfies Eqs.~\eqref{Eq:Cond1} and \eqref{Eq:Cond3}. Secondly, for the mass ratio of $\frac{m}{M}=6\times10^{-13}$ and an amplitude reflectivity $r=0.45$ we choose a ratio $\frac{\delta}{\gamma}\simeq 450$ in order to approximately satisfy condition \eqref{Eq:Cond2} and at the same time to ease requirements for condition \eqref{Eq:Cond4}. Thirdly, from the data measured in \cite{Zink2004} we infer a value of $k_B\kappa_{th}\simeq 10~$nW/K for the dimensions of the membrane $(100\,\mu{\rm m}\times 100\,\mu{\rm m}\times 50\,{\rm nm})=(l\times l\times d)$ required here \footnote{In \cite{Zink2004} a thermal link of $k_B\kappa'_{th}=0.1\,\mu$W/K was measured for a thin, square membrane $(d'=200\,{\rm nm},\,l'=5\,{\rm mm})$ with power dissipated in a central square area ($l'_1=2.5\,$mm). From the solution of the Laplace equation we estimate the thermal link to scale like $\frac{\kappa_{th}}{\kappa'_{th}}\simeq \frac{d}{d'}\ln\big(\frac{l'}{l'_1}\big)/\ln\big(\frac{l}{2 w_0}\big)\simeq 0.1$. This is confirmed by a finite-element simulation.}. A mechanical quality factor of $Q=10^7$ and a resonance frequency $\omega_m=2\pi\times 1.3~$MHz set the left hand side of Eq.~\eqref{Eq:Cond4} to $\sim45$. Finally, the resonance condition $\omega_{at}=\omega_m$ demands a circulating power $P_c\simeq 850\,\mu$W which will cause heating of $\lesssim2.5$\,K for the given thermal link. We assume the driving laser to be shot noise limited in intensity at the relevant sideband frequencies $\omega_m,\,2\omega_m$ and to have kHz linewidth, in order to avoid FM to AM conversion of frequency noise \cite{Miller2005}. This is readily achieved at the optical frequencies and $\mu$W driving power required here. In order to make a statement about the absolute timescales of the dynamics, we still need to fix the cavity length. For $L=50\,\mu$m we find a cavity mediated coupling $G\simeq2\pi\times45\,$kHz and decoherence rates $\Gamma_c,\,\Gamma_m,\,\Gamma_{at}\simeq 0.1\times G  $. It it thus indeed possible to enter the strong coupling regime with state of the art experimental parameters.

\begin{figure}[t]
\includegraphics[width=0.95\columnwidth]{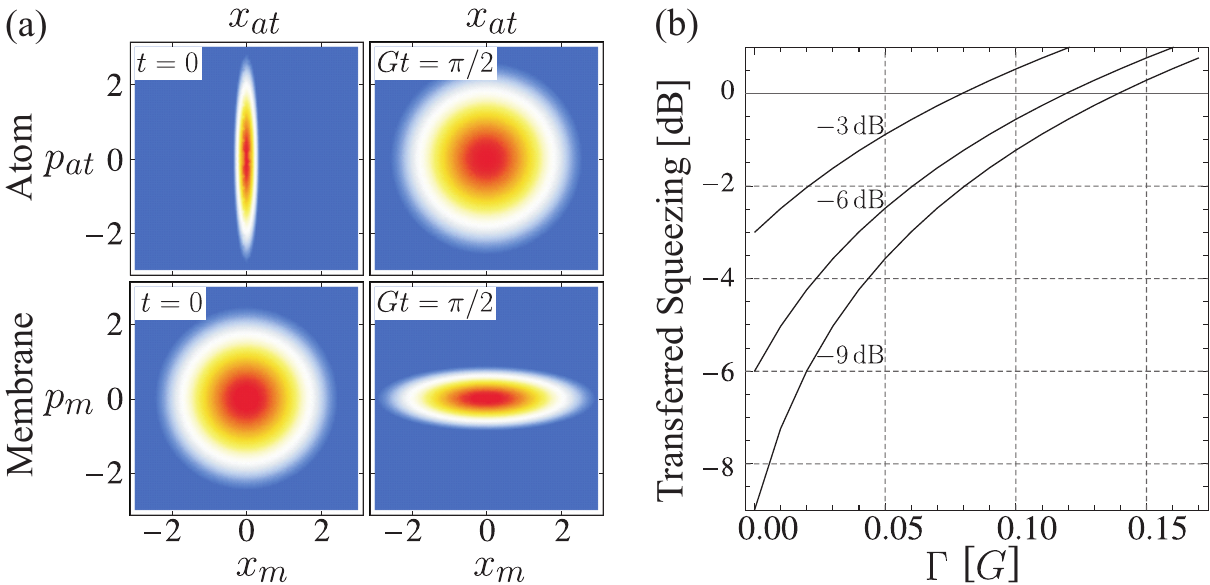}
\caption{(a) Wigner functions of atom and membrane (upper and lower panels, respectively). At $t=0$ (left panels) the atom is in a squeezed state (9\,dB) and the membrane in a thermal state with a mean number of phonons $\bar{n}=5$. An exact solution of the equation of motion \eqref{Eq:EffMEQ} with losses $\Gamma=\Gamma_c,\Gamma_m,\Gamma_{at}$ at rate $\Gamma=0.1\times G$ shows that after a time $Gt=\frac{\pi}{2}$ (right panels) the states are exchanged, up to a trivial rotation in phase space by $90^\circ$. (b) Squeezing transferred to membrane (maximized over time), versus loss rate $\Gamma$, for the indicated values of initial atomic position fluctuations.}
\label{Fig:Transfer}
\end{figure}

While being a surprising result on its own, entering the regime of strong coupling holds promise for diverse applications, including for preparation and readout of quantum states of mesoscopic massive oscillators. In the regime $\omega_m=\omega_{at}\gg G$, where the rotating wave approximation can be applied in Eq.~\eqref{Eq:Heff}, the effective dynamics is described by $H_I\simeq G(a_m a^\dagger_{at}+{\rm h.c.})$ in the interaction picture. This interaction swaps the state of the atom and the membrane after a time $Gt=\frac{\pi}{2}$. Thus, states which are easily created on the side of the atom (e.g.,~squeezed or Fock states) can be transferred to the membrane. In Fig.~\ref{Fig:Transfer} we study such a transfer of a squeezed state based on the exact solution of the master equation in Eq.~\eqref{Eq:EffMEQ}. The figure also illustrates the importance of limiting the loss in order to achieve quantum state transfer or readout. The general analysis provided here shows that condition \eqref{Eq:Cond4} is the principal bottleneck for a reduction of losses. Especially the ratio $\frac{\kappa_{th}\mathcal{F}^2}{\gamma_mM}$ might be further increased by improving material properties and nanostructuring, though there will always be an apparent tradeoff between good mechanical isolation and a large thermal link. Another rather obvious route for improvement is to use a small ensemble of $N$ atoms trapped inside the cavity \cite{Colombe2007,Murch2008,Brennecke2008}, resulting in a $\sqrt{N}$ enhancement of the atom-cavity coupling. However, our main point here is to identify the general conditions for achieving strong coupling of a single atom to a massive mechanical oscillator, and to demonstrate that it is possible to meet these conditions with state of the art systems.

Support by the Austrian Science Fund through SFB FOQUS, by the Institute for Quantum Optics and Quantum Information, by the European Union through project EuroSQIP, by NIST and NSF, and by the DFG through NIM, SFB631 and the Emmy-Noether program is acknowledged. MW, KH, PZ and JY thank HJK for hospitality at Caltech.



\end{document}